\documentclass{amsart}
\usepackage{graphicx}
\usepackage{amsfonts}
\begin{document}

\title{Variations on the Fibonacci Universal Code}

\author{James Harold Thomas}
\address{Department of Electrical and Computer Engineering\\
Louisiana State University\\
Baton Rouge, Louisiana, 70803, USA} 

\begin{abstract}
This note presents variations on the Fibonacci universal code, that may also  be called the Gopala-Hemachandra code, that can have applications in source coding as well as in cryptography.
\end{abstract}

\maketitle

\section{Introduction}

For the purposes of Fibonacci coding, the Fibonacci sequence of order two, $F(k)$, where $k \in \mathbb{N},\,i>0$ is defined as ~\cite {apostolico},
\[
F(0)=1,\,F(1)=2,\,F(k)=F(k-1)+F(k-2),\,\forall k>2.
\]
We will consider this familiar Fibonacci sequence as well as alternative sequences with different initial values.

The Fibonacci code is a type of universal coding scheme that maps the positive integers, which represent the probability rank of source messages, into variable length codewords. The codeword elements have a binary alphabet set, and are defined according to the following rule for a given positive integer $n$. Construct a vector $A(n)$ of Fibonacci numbers such that the $i$th element of $A(n)$, $A(n)_{i}=F(i),\,i=0,1,\dots,d$, where $F(d)$ is the largest Fibonacci number less than or equal to $n$. A second vector $B(n)$ of binary digits is then chosen, also with dimension $d$, such that $A(n)^{T} \cdot B(n)=n$, and $B(n)_{d}=1$. The codeword, $FB(n)$, is the vector with dimension $d+1$ where $FB(n)_{k}=B(n)_{k}$ for $1 \le k \le d$, and $FB(n)_{d+1}=1$.

For example, suppose we wish to construct a codeword for the integer 10. Since F(4)=8 is the largest Fibonacci number less than or equal to 10, then $d=4$ in this example. The vectors $A(n)$, $B(n)$, and $FB(n)$ are given as
\[
	A(n) = 
	\begin{pmatrix}
		1\\2\\3\\5\\8
	\end{pmatrix}, \,
	B(n) = 
	\begin{pmatrix}
		0\\1\\0\\0\\1
	\end{pmatrix}, \,
	FB(n) = 
	\begin{pmatrix}
		0\\1\\0\\0\\1\\1
	\end{pmatrix}
\]

Zeckendorf's theorem states that every positive integer has a unique representation as the sum of nonconsecutive Fibonacci numbers ~\cite {Zeckendorf}. An integer written in such a fashion is said to be in Zeckendorf representation. Therefore while the recursive nature of the Fibonacci numbers allow some integers to have multiple representations using the above scheme, we can always choose $B(n)$ such that there are no two consecutive $1$'s. For example, the decimal number 10 can be represented as $F(1) + F(2) + F(3)$ by $B(10) = \begin{pmatrix} 0111 \end{pmatrix}^{T}$ or in Zeckendorf representation as $F(1) + F(4)$ by $B(10) = \begin{pmatrix} 01001 \end{pmatrix}^{T}$. Since $FB(10)_{d+1} = FB(10)_{d} = 1$, the only instance of two consecutive $1$'s in the codeword $FB(n)$ is at its termination when we choose the Zeckendorf representation, thus giving the code the prefix condition.

\begin{table}[h]
\begin{tabular}{|l|l|}
\hline
$n$&$FB(n)^{T}$\\
\hline

1&
$\begin{pmatrix} 11 \end{pmatrix}$\\

2&
$\begin{pmatrix} 011 \end{pmatrix}$\\

3&
$\begin{pmatrix} 0011 \end{pmatrix}$\\

4&
$\begin{pmatrix} 1011 \end{pmatrix}$\\

5&
$\begin{pmatrix} 00011 \end{pmatrix}$\\

6&
$\begin{pmatrix} 10011 \end{pmatrix}$\\

7&
$\begin{pmatrix} 01011 \end{pmatrix}$\\

8&
$\begin{pmatrix} 000011 \end{pmatrix}$\\

9&
$\begin{pmatrix} 100011 \end{pmatrix}$\\

10&
$\begin{pmatrix} 010011 \end{pmatrix}$\\

11&
$\begin{pmatrix} 001011 \end{pmatrix}$\\

12&
$\begin{pmatrix} 101011 \end{pmatrix}$\\

13&
$\begin{pmatrix} 0000011 \end{pmatrix}$\\

14&
$\begin{pmatrix} 1000011 \end{pmatrix}$\\

15&
$\begin{pmatrix} 0100011 \end{pmatrix}$\\
\hline
\end{tabular}
\caption{The Fibonacci Code, $n = 1, \dots , 15$}
\label{tab:1}
\end{table}

\section{Gopala-Hemachandra Sequence and Codes}

A variation to the Fibonacci sequence is the more general Gopala-Hemachandra sequence ~\cite {kak1}: 

\[
\{a, \, b, \, a+b, \, a+2b, \, 2a+3b, \, 3a+5b, \dots \},
\]
for any pair $a$, $b$, which for the case $a=1$, $b=2$ represents the Fibonacci numbers. For a historical context of these sequences, see ~\cite {kak2,kak3,pearce}.
	
We now introduce a variation on the Fibonacci coding scheme by using the Gopala-Hemachandra sequence to construct $A(n)$. Define a second order Variant Fibonacci sequence, $VF_{a}(n)$, as the Gopala-Hemachandra sequence above such that $b=1-a$. That is, $VF_{a}(0)=a, (a\in \mathbb{Z})$, $VF_{a}(1)=1-a$, and for $n \geq 2$, $VF_{a}(n)=VF_{a}(n - 1)+VF_{a}(n-2)$. With this definition we obtain (e.g.) $VF_{-2}(n)$ as $\{-2,\,3,\,1,\,4,\,5,\,9,\,14,\,23,\dots\}$. We note in passing that similar Variant Fibonacci sequences have been investigated in the construction of hypercubes ~\cite {wu}.

While the term ``Zeckendorf representation'' is properly used only in reference to the standard Fibonacci sequence, we will use it when discussing similar representations of numbers based on Variant Fibonacci sequences. Daykin proved that only the standard Fibonacci sequence $F(n)$ gives all positive integers a unique Zeckendorf representation ~\cite {daykin}. Thus the Variant Fibonacci sequences allow for multiple Zeckendorf representations of the same integer. With these Variant Fibonacci sequences, we can obtain a new universal source code, which we call the Gopala-Hemachandra (G-H) code, using the same rule used to generate the standard Fibonacci code $FB(n)$ above.
	
We quickly discover that not all values of $a$ will generate a Variant Fibonacci sequence which is suitable for all applications of universal coding. For example, for $a=-5$, we obtain 
$VF_{-5}(n)=\{-5,\,6,\,1,\,7,\,8,\,15,\,23,\,38,\dots\}$. It is easily seen that there is no Zeckendorf representation of the integers $5$ or $12$ using this sequence.

\begin{table}[h]
\begin{tabular}{|l|l|l|l||l|}
\hline
$n$&$GH_{-2}(n)$&$GH_{-3}(n)$&$GH_{-4}(n)$&$GH_{-5}(n)$\\
\hline

1&
$\begin{pmatrix} 0011 \end{pmatrix}$&
$\begin{pmatrix} 0011 \end{pmatrix}$&
$\begin{pmatrix} 0011 \end{pmatrix}$&
$\begin{pmatrix} 0011 \end{pmatrix}$\\

2&
$\begin{pmatrix} 10011 \end{pmatrix}$&
$\begin{pmatrix} 10011 \end{pmatrix}$&
$\begin{pmatrix} 10011 \end{pmatrix}$&
$\begin{pmatrix} 10011 \end{pmatrix}$\\

3&
$\begin{pmatrix} 011 \end{pmatrix}$ or $\begin{pmatrix} 100011 \end{pmatrix}$&
$\begin{pmatrix} 100011 \end{pmatrix}$&
$\begin{pmatrix} 100011 \end{pmatrix}$&
$\begin{pmatrix} 100011 \end{pmatrix}$\\

4&
$\begin{pmatrix} 00011 \end{pmatrix}$ or $\begin{pmatrix} 101011 \end{pmatrix}$&
$\begin{pmatrix} 011 \end{pmatrix}$ or $\begin{pmatrix} 101011 \end{pmatrix}$&
$\begin{pmatrix} 101011 \end{pmatrix}$&
$\begin{pmatrix} 101011 \end{pmatrix}$\\

5&
$\begin{pmatrix} 000011 \end{pmatrix}$&
$\begin{pmatrix} 00011 \end{pmatrix}$&
$\begin{pmatrix} 011 \end{pmatrix}$&
N/A\\

6&
$\begin{pmatrix} 001011 \end{pmatrix}$&
$\begin{pmatrix} 000011 \end{pmatrix}$&
$\begin{pmatrix} 00011 \end{pmatrix}$&
$\begin{pmatrix} 011 \end{pmatrix}$\\

7&
$\begin{pmatrix} 01011 \end{pmatrix}$ or $\begin{pmatrix} 1000011 \end{pmatrix}$&
$\begin{pmatrix} 001011 \end{pmatrix}$&
$\begin{pmatrix} 000011 \end{pmatrix}$&
$\begin{pmatrix} 00011 \end{pmatrix}$\\

8&
$\begin{pmatrix} 010011 \end{pmatrix}$ or $\begin{pmatrix} 1010011 \end{pmatrix}$&
$\begin{pmatrix} 1000011 \end{pmatrix}$&
$\begin{pmatrix} 001011 \end{pmatrix}$&
$\begin{pmatrix} 000011 \end{pmatrix}$\\

9&
$\begin{pmatrix} 0000011 \end{pmatrix}$&
$\begin{pmatrix} 01011 \end{pmatrix}$ or $\begin{pmatrix} 1010011 \end{pmatrix}$&
$\begin{pmatrix} 1000011 \end{pmatrix}$&
$\begin{pmatrix} 001011 \end{pmatrix}$\\

10&
$\begin{pmatrix} 0010011 \end{pmatrix}$&
$\begin{pmatrix} 010011 \end{pmatrix}$&
$\begin{pmatrix} 1010011 \end{pmatrix}$&
$\begin{pmatrix} 1000011 \end{pmatrix}$\\

11&
$\begin{pmatrix} 1001011 \end{pmatrix}$&
$\begin{pmatrix} 0000011 \end{pmatrix}$&
$\begin{pmatrix} 01011 \end{pmatrix}$&
$\begin{pmatrix} 1010011 \end{pmatrix}$\\

12&
$\begin{pmatrix} 0100011 \end{pmatrix}$ or $\begin{pmatrix} 10000011 \end{pmatrix}$&
$\begin{pmatrix} 0010011 \end{pmatrix}$&
$\begin{pmatrix} 010011 \end{pmatrix}$&
N/A\\

13&
$\begin{pmatrix} 0001011 \end{pmatrix}$ or $\begin{pmatrix} 10100011 \end{pmatrix}$&
$\begin{pmatrix} 1001011 \end{pmatrix}$&
$\begin{pmatrix} 0000011 \end{pmatrix}$&
$\begin{pmatrix} 01011 \end{pmatrix}$\\

14&
$\begin{pmatrix} 00000011 \end{pmatrix}$&
$\begin{pmatrix} 10000011 \end{pmatrix}$&
$\begin{pmatrix} 0010011 \end{pmatrix}$&
$\begin{pmatrix} 010011 \end{pmatrix}$\\

15&
$\begin{pmatrix} 00100011 \end{pmatrix}$&
$\begin{pmatrix} 0100011 \end{pmatrix}$ or $\begin{pmatrix} 10100011 \end{pmatrix}$&
$\begin{pmatrix} 1001011 \end{pmatrix}$&
$\begin{pmatrix} 0000011 \end{pmatrix}$\\

\hline
\end{tabular}
\caption{Some G-H Codes, $n = 1, \dots , 15$}
\label{tab:2}
\end{table}

Even though the G-H codes are longer than the standard Fibonacci code and therefore would less desirable by themselves, the family of G-H codes, $GH_{a}(n)$, which satisfy our condition for encoding all of the desired integers $1 \le n \le M$ (where $M$ is the number of possible source messages), allows us to have many more universal codes at our disposal when transmitting a message. Even those G-H codes such as $GH_{-5}(n)$, which lack the ability to encode certain positive integers, could be used on portions of a message signal that contained only those source messages which they are able to encode.

\section{Conclusions}
The encoding scheme outlined in this paper offers some highly desirable cryptographic properties. Since the G-H codes are uniquely determined by their initial value $a$, the codebook could be easily changed multiple times during transmission, making decoding much more difficult. In addition the presence of multiple representations of the same integer allow for a codebook that appears larger than it actually is. As shown in Table \ref{tab:2}, we also see that the codeword lengths are not always increasing. This property seems to be undesirable by itself, but it could also offer cryptographic advantages.

\section*{Acknowledgements}
The author thanks the Louisiana Board of Regents, BoRSF, under agreement NASA/LEQSF(2005-2010)-LaSPACE and NASA/LaSPACE under grant NNG05GH22H for support during this project.

\end{document}